\documentclass[11pt]{article}

\usepackage{algorithm}
\usepackage{algorithmic}
\usepackage{tabularx}
\usepackage{url}
\usepackage{graphicx}
\usepackage{amsmath}

\title{Duplicate Detection with Efficient Language Models for Automatic Bibliographic Heterogeneous Data Integration }
\author{Nicolas Turenne\thanks{nturenne.inra@yahoo.fr}}
\date{}

\begin{document}

\maketitle 

\begin{center}
Univ. Paris-Est, LISIS, INRA \\
F-77450 Champs-sur-Marne, France \\
\end{center}

\begin{abstract}
   We present a new method to detect duplicates used to merge different bibliographic record corpora with the help of lexical and social information. As we show, a trivial key is not available to delete useless documents. Merging heteregeneous document databases to get a maximum of information can be of interest. In our case we try to build a document corpus about the TOR molecule so as to extract relationships with other gene components from PubMed and WebOfScience document databases. Our approach makes key fingerprints based on n-grams. We made two documents gold standards using this corpus to make an evaluation. Comparison with other well-known methods in deduplication gives best scores of recall (95\%) and precision (100\%).  
~

\end{abstract}









\section[Introduction]{Introduction}\label{sec:int}
Since early 90ies with internet and low-cost workstation development, lots of large public or commercial databases emerged. Sometimes fusion of data from different databases seemes to be a good strategy to build a convenient document repository about a specific topic. In this way cleaning and normalization is necessary. If deduplication between documents from a same data repository is trivial (through title or identifier matching), indexing over databases is not the same and just comparing whole title is not sufficient (see below for examples). In this paper we are interested in deduplication (also called dedupes) between two specifc databases WebOfScience (commercial) and PubMed (public). There are two tasks for such data processing. First, it can be understood as the elimination of redundant data in computer storage (see \cite{Yampolskii:1973}, \cite{Hickey:1979}, \cite{Bitton:1983}, \cite{Goyal:1987}]). Second, it is related to record linkage, in databases: that is, finding entries that refer to the same entity in two or more files (see  \cite{Newcombe:1959},\cite{Fellegi:1969}). Most duplicate detection methods can be classified as:  
\begin{enumerate}
	\item domain-dependent: requiring some knowledge of the data source. For instance, data fields. 
  \item domain-independent: not requiring any knowledge of the data source. Generally this uses text strings within the records to search for matches. 
\end{enumerate}
Our goal is to merge records from bibliographic documents not knowing the structure. Some approaches try to detect near-duplicates aiming at identifying plagiarism (see  \cite{Hoad:2003},\cite{Deng:2006},\cite{gong:2008},\cite{Errami:2008}). Others focus, as our approach, on detecting exact duplicates, so as to merge or delete entries when merging items between heteregeneous databases (see  \cite{Ridley:1992},\cite{Madhavaram:1996},\cite{Monge:1997},\cite{Yager:1998},\cite{Paskalev:2006},\cite{Muthmann:2009}).  \hspace*{\fill} \\
If a primary key exists, the duplicate records could be easily identified. Such a key exists for a document database and is called a DOI (Document Object Identifier). It looks like 
\begin{center}
\textit{10.1016/j.bcp.2010.01.019}. 
\end{center}
Its use leads to visualization of the document's electronic version on the Internet. But the use of such a key is recent and building useful corpora including documents published before 2005 makes this approach irrelevant.  \\
Comparing two documents can be, in the case of short texts, a string alignment. Some works have lead to interesting processing techniques such as the Levenshtein (or edit) distance  \cite{Levenshtein:1966}, based on a cost computation for the deletion and insertion of characters to transform one string into another; or fast algorithms in the tradition of \cite{Knuth:1977} and \cite{Boyer:1977}, in the area of DNA sequences alignment these are well described by \cite{Smith:1981} and \cite{Altschul:1997}. All current techniques for solving the duplicate discovery problem in a document collection are based on document fingerprinting, in which a compact representation of a selected subset of contiguous text chunks occurring in each document-its fingerprint-is stored. Pairs of documents are considered as possible duplicates if enough of the chunks in their respective fingerprints are matched. A chunk is defined \cite{Broder:1997}, as a contiguous subsequence; that is, a chunk represents a contiguous set of words or characters within the document.  \hspace*{\fill} \\
Section~\ref{sec:bidc} presents the data and our application in molecular biology. In Section~\ref{sec:sadsd}, we present the state of the art of deduplication and some definitions. In Section~\ref{sec:olmf} we introduce our language model methodology of fingerprinting. Finally in Section~\ref{sec:ev} we show the results of evaluation and comparison with six other approaches: three fingerprinting approaches (simple fields-key method, multi-fields-key method, anchor method), two scoring approaches (Salton information retrieval method, collocation method), and the random approach (Monte Carlo method).

\section[Biological issue and document collections]{The biological issue and document collections}\label{sec:bidc}

In this section, we present how and for what the document collections were built. 

\subsection[The biological issue]{The biological issue}\label{subsec:bi}

The background issue concerns understanding how the TOR molecule is involved in terms of upper and down regulation of other genes in the framework of a species (such as the plant Arabidopsis thaliana). TOR gets lots of interest in the research community because of its role in growth and metabolism. Comparison with other species is a knowledge tool often used by biologists. TOR has similar genes (called orthologs) to other species (mouse, human, yeast, worm and fly). A corpus covering all possible links between TOR and other molecules (TOR-interactomes) could be of great interest, and secondly, the interaction between components of two TOR-interactomes could also brings some pieces of knowledge. It should be interesting to detect equivalent genes using existing databases and the BLAST approach (DNA sequence comparison) but databases are not always well annotated. Usually the NCBI pubmed is viewed by biologists to search documents and even biological data. Many others sources of documents are widespread. A famous one is Web of Science. It claims encapsulation of pubmed but updates are not running well. Our idea for getting a larger corpus of documents relies on the fusion of data collected from different databases to permit the maximum extraction of relevant relationships between gene associations.

\subsection[Document collections (corpora)]{Document collections (corpora)}\label{subsec:dcc}

We used for convenience two well known databases to build our corpora. The first one is PubMed, indexing more than 19 million documents with an update of 3,000 documents/day. And the Web of Science indexing more than 90 millions documents with an update of 5000 documents per day. Since a couple of years ago, Web of Science has been integrating PubMed.

A manually defined query based on TOR variants names has been applied to both databases (PubMed, hereafter PM, and the Web of Science, hereafter WOS).

query:
(raptor[Title/Abstract] OR kog1[Title/Abstract] OR lst8[Title/Abstract] OR "target \hspace*{\fill} \\ of rapamycin"[Title/Abstract] OR TOR[Title/Abstract] OR TORC1[Title/Abstract] OR \hspace*{\fill} \\ TORC2[Title/Abstract] OR mTORC1[Title/Abstract] OR Dd-TOR[Title/Abstract] OR \hspace*{\fill} \\
mTORC2[Title/Abstract] OR TOR1[Title/Abstract] OR TOR2[Title/Abstract] OR \hspace*{\fill} \\ mTOR[Title/Abstract] OR dTOR[Title/Abstract] OR CeTOR[Title/Abstract] OR \hspace*{\fill} \\ AtTOR[Title/Abstract] OR Tor1p[Title/Abstract] or Tor2p[Title/Abstract]) not "tor vergata" \hspace*{\fill} \\

The result of the query is then cleaned by excluding documents from the following topics.

Refined by: [excluding] Subject Areas=( Meteorology \& Atmospheric Sciences OR Telecommunications  OR Education \& Educational Research OR Astronomy \& Astrophysics OR Instruments \& Instrumentation OR Nursing OR Computer Science OR Social Issues OR Communication OR Radiology, Nuclear Medicine \& Medical Imaging OR Water Resources OR Optics OR Nuclear Science \& Technology OR History OR Information Science \& Library Science OR Geochemistry \& Geophysics OR Energy \& Fuels OR Polymer Science OR Mathematics OR Business \& Economics OR Government \& Law OR Geography OR Sport  Sciences OR Anthropology OR Sociology OR Engineering OR Mechanics OR Materials Science OR Automation \& Control Systems OR Physics OR Family Studies OR Psychology OR Geology OR Imaging  Science \& Photographics Technology )

\begin{figure}[!ht]
\centering
\renewcommand{\arraystretch}{1.2}
{\scriptsize
\begin{tabularx}{15.5cm}{|>{\centering\hsize=0.4\hsize\arraybackslash}X|>{\centering\hsize=0.7\hsize\arraybackslash}X|>{\centering\hsize=0.7\hsize\arraybackslash}X|>{\centering\hsize=0.5\hsize\arraybackslash}X|>{\centering\hsize=0.6\hsize\arraybackslash}X|>{\centering\hsize=0.8\hsize\arraybackslash}X|>{\centering\arraybackslash}X|>{\centering\hsize=0.5\hsize\arraybackslash}X|}
   \hline
   \textbf{  } & \textbf{ \#docs with filtering } & \textbf{ \#docs} & \textbf{ \#words } & \textbf{ year range } & \textbf{ \#Field Author} & \textbf{\#Field Title} & \textbf{\#Field Source} \\
   \hline
   \textbf{PM} & 7,709 & 7,709 & 3,870,000 & 1951-2010 & 7,709 & 7,689 & 7,709 \\	 	 	 
   \hline
   \textbf{WOS} & 12,658 & 16,178 & 2,780,000 & 1951-2010 & 12,658 & 12,658 & 12,658 \\
   \hline
\end{tabularx} } 
\caption{Statistics about corpora content.}\label{figure1}
\end{figure}

An Overview of corpora size and number of some important fields are shown on Figure  \ref{figure1}.

\subsection[Duplication issue]{Duplication issue}\label{subsec:di}

As mentionned in the Introduction, when a unique identifier is attached to a document, its usage is immediate to avoid any duplicated insertion during a database fusion operation.

Such an identfier does not systematically exist for all documents and moreover we observe that some differences for the same document can occur in the title.

For instance the sentence below belongs to the same publication but is indexed differently in PM and WOS databases:

\begin{enumerate}
	\item PM sentence: \textit{Here we show that CRP is required for the biosynthesis of cholera autoinducer 1 (CAI-1)  (PMID-17768239)}
	\item 'WOS' sentence: \textit{Here we show that CRP is required for the biosynthesis of cholera autoinducer 1 (CAl-1) (UT WOS:000250013400014)}
\end{enumerate}

They contain a different writing of the gene name cai-1.

This is another example:  
\begin{enumerate}
	\item WOS sentence : \textit{Structural Analysis and Functional Implications of the Negative mTORCl Regulator REDD1} (UT WOS:000275711400021)
	\item PM sentence : \textit{Structural analysis and functional implications of the negative mTORC1 regulator REDD1.} (PMID- 20166753)
\end{enumerate}
In this case the gene name mTorc1 is written with l in PM instead of 1 in the WOS

A last example, insertion or permutation of characters can occur as in :
\begin{enumerate}
	\item WOS sentence :  \textit{Bis(morpholino-1,3,5-triazine) Derivatives: Potent Adenosine 5} \\
   \textit{'-Triphosphate Competitive Phosphatidylinositol-3-kinase/Mammalian} \\
   \textit{Target of Rapamycin Inhibitors: Discovery of Compound 26 (PKI-587), a}\\
   \textit{Highly Efficacious Dual Inhibitor}  (UT WOS:000275805900058)
	\item PM sentence : \textit{Bis(morpholino-1,3,5-triazine) derivatives: potent adenosine 5'-triphosphate}\\
      \textit{competitive phosphatidylinositol-3-kinase/mammalian target of } \\
      \textit{rapamycin inhibitors: discovery of compound 26 (PKI-587), a highly efficacious dual} \\
      \textit{inhibitor.}  (PMID- 20188552)
\end{enumerate}
where 5'-Triphosphate is cut after 5 in the WOS document. When reconstructing the sentence, the cut is replace by a blank and makes matching impossible.

The duplication issue is a two-class problem where classes are 'duplicate' tag and 'non-duplicate' tag. A test-database is processed by an algorithm. Such algorithm pick up each document from the test-database to transform it and compare to each document, transformed in the same way, of a target-database. For our purpose, the test-database is the PM database and the target database is the WOS database. 
\section[State of the art of deduplication and some definitions]{state of the art of deduplication and some definitions}\label{sec:sadsd}

In this section, we present some definitions about linguistic parsing and the best algorithms for deduplication.

\subsection[Definitions]{Definitions about formal content of a document}\label{subsec:dfcd}

The approach we adopt to develop, usually concerning the scientific community, relies on a definition of a language model to compare document contents.

Firstly we need to define what is an elementary alphabet.

\newtheorem{thm}{Theorem}[section]
\newtheorem{definition}[thm]{Definition}

\begin{definition} \textup{ Alphabet  }\hspace*{\fill} \\
\textup{ Let A be the set of characters [a,b,c,...,z;0,1,...,9] to make a word; some special characters can belong to a word sucha as [-']. }
\end{definition} 

Content description requires some delimiters:

\begin{definition} \textup{ Delimiter  }\hspace*{\fill} \\
\textup{ Let D be the set of special characters $[space=_.;:{}()*+.,?\!\%[]"]$; the $space$ character is the main delimiter. }
\end{definition} 

Then we can define general string.

\begin{definition} \textup{ String  }\hspace*{\fill} \\
\textup{ A string S, of length N, is a set of words $\{W_{i}\}_{i=1,N}$ separated by one or several delimiter(s). }
\end{definition} 

At the word level, interesting patterns in a string could be elementary sequences we call a n-gram:

\begin{definition} \textup{ n-gram  }\hspace*{\fill} \\
\textup{ A n-gram is a set of n consecutive character(s) $\{C_{j}\}_{i=1,n}$ in a string where $C_{j}$ belongs to A. }
\end{definition} 

In documents, useful components are described by associations. These associations, we call n-collocations, defines semantic links:

\begin{definition} \textup{ n-collocation (or chunk)  }\hspace*{\fill} \\
\textup{ An n-collocation is a string of n consecutive words  $\{W_{i}\}_{i=1,n}$ . }
\end{definition} 

Finally at the document scale, a document is divided into several fields having different kinds of descriptors such as dates, authors, titles, sources of publication for instance:

\begin{definition} \textup{ Fields  }\hspace*{\fill} \\
\textup{ A bibliographic document is composed of a set of fields  $F_{k}$ ; each $F_{k}$ begins with an identifier and is followed by a string.  }
\end{definition} 

Some algorithms requires extraction of a key:
\begin{definition} \textup{ Key  }\hspace*{\fill} \\
\textup{ A Key K attached to a given document is a language model, defined by the alphabet and/or strings and/or collocations and/or n-grams included in some fields of the document. K can be built for any document contained in a document database, and K is supposed to be unique.  }
\end{definition} 

\subsection[Main approaches]{Main approaches}\label{subsec:ma}

Two families are well applied for deduplication: fingerprint approaches and similarity-based approaches  (see \cite{Elmagarmid:2007}, \cite{Potthast:2008}).

Let us consider, to begin with, the following fingerprint approaches.

\cite{Hernandez:1995} proposed the sorted neighbourhood. This is done in three phases.

\algsetup{indent=2em}
\newcommand{\Rc}{\ensuremath{\mbox{\sc Rc}}}

\begin{algorithm}[H]
\caption{Sorted Neighborhood (SN) algorithm}\label{alg:SN}
\begin{algorithmic}[1]
\medskip

\STATE $Create\ a\ Key\ for\ each\ record$
\STATE $Sort\ records\ on\ this\ key$
\STATE $Merge/Purge\ records$

\end{algorithmic}
\end{algorithm}

Considering our purpose for comparing documents, it is convenient to adapt and derive some interesting variants of "Algorithm~\ref{alg:SN}" using some fields. Some specific fields can serve directly or be combined as a key to make a fingerprint. A first variant, we will call it author-fingerprint (AF), make the key with the name of the first author. A second variant can be the selection of the title as a key, we call it title-fingerprint (TF). Based on the previous we can lightly process the title by deleting a delimiter such as space, we call this variant modified-title-fingerprint (MTF). Finally we can imagine, knowing what type of document is a scientific document, that meaningful combined fields should be first author, date and source of publication, so the key would be the concatenation of these fields; we call this approach author-review-date fingerprint (ARDF).

Let suppose the following example :

Let consider a real document published with the following fields (Authors, Title, Source of publication and Date):  
\begin{itemize}
	\item AU  -  Ayral-Kaloustian, S     Gu, JX     Lucas, J     Cinque, M    Gaydos, C    Zask, A    Chaudhary, I    Wang, JY    Di, L    Young, M    Ruppen, M    Mansour, TS    Gibbons, JJ    Yu, K
	\item TI  - Hybrid Inhibitors of Phosphatidylinositol 3-Kinase (PI3K) and the
   Mammalian Target of Rapamycin (mTOR): Design, Synthesis, and Superior
   Antitumor Activity of Novel Wortmannin-Rapamycin Conjugates
	\item SO  - JOURNAL OF MEDICINAL CHEMISTRY
	\item DP  - 2010
\end{itemize}

The trivial keys (K) with the previous methodologies should be, after transforming the strings to lower case:  
\begin{enumerate}
	\item $K_{AF}$ =  \textit{ayral-kaloustian}
	\item $K_{TF}$ = \textit{hybrid inhibitors of phosphatidylinositol 3-pinase (pi3k) and the mammalian target of rapamycin (mtor): design, synthesis, and superior antitumor activity of novel wortmannin-rapamycin conjugates}
	\item $K_{MTF}$ = \textit{hybridinhibitorsofphosphatidylinositol3-pinase(pi3k)andthemammaliantargetof rapamycin(mtor):design,synthesis,andsuperiorantitumoractivityofnovelwortmannin-rapa \\ mycinconjugates}
	\item $K_{ARDF}$ = \textit{ayral-kaloustian-journal of medicinal chemistry-2010}
\end{enumerate}

Sort data has time complexity $O(NlogN)$ for a good algorithm, $O(N^2)$ for a bad algorithm. To optimize at step 3 of "Algorithm~\ref{alg:SN}", move a fixed size window through the sequential list of records. This limits comparisons to the records in the window. The multi-pass variant (SNM) involves using multiple passes of the initial SN.

\cite{Feekin:2000} developped the k-way sorting method for no uniquely distinguishing data fields.
The concept behind the k-way sort method follows (see "Algorithm~\ref{alg:k-way}").

\begin{algorithm}[H]
\caption{k-way algorithm}\label{alg:k-way}
\begin{algorithmic}[1]
\medskip

\STATE $Let\ k\ be\ number\ of\ columns\ to\ be\ used\ for\ sorting$
\STATE $Select\ the\ k\ most\ meaningful\ combination$
\STATE $Assign\ a\ record\ identifier\ to\ each\ record$
\STATE $Sort\ records$
\STATE $Compare\ adjacent\ rows:$
	\IF {$matched\ columns\ >\ k/2\ columns$}
		\STATE $the\ two\ records\ match$
	\ENDIF
\end{algorithmic}
\end{algorithm}

(see "Algorithm~\ref{alg:GF}") describes basic fingerprinting process is as follows (see \cite{Bernstein:2004}, \cite{Shivakumar:1995}, \cite{Manber:1994}):

\begin{algorithm}[H]
\caption{General fingerprinting algorithm}\label{alg:GF}
\begin{algorithmic}[1]
\medskip
\STATE $documents\ in\ a\ collection\ are\ parsed\ into\ units\ (typically\ either\  $
$characters\ or\ individual\ words)$
\STATE $representative\ n-collocations\ are\ selected\ through\ the\ use\ of\ a\ heuristic$
\STATE $the\ selected\ chunks\ are\ then\ hashed\ for\ efficient\ retrieval\ and/or\ compact\ storage$
\STATE $the\ hash\ keys,\ and\ possibly\ also\ the\ chunks\ themselves,\ are\ then\ stored,\ often\ in\ a\   $ 
$inverted\ index\ structure$
\STATE $Compare\ co-derivatives\ documents\ using\ index\ of\ hash\ keys$
\end{algorithmic}
\end{algorithm}

The principal way in which algorithms differentiate themselves is in the choice of selection heuristic, the method of determination which chunks should be selected for storage in each document's fingerprint.

\cite{Hylton:1996} envisioned an entity grouping approach for identifying author-title clusters in bibliographical records.

\begin{algorithm}[H]
\caption{author-title clustering algorithm}\label{alg:AT}
\begin{algorithmic}[1]
\medskip
\STATE $query\ of\ first\ author\ name\ and\ two\ words\ of\ the\ title\ gathering\ a\ pool\ of\ documents\ $
$repeated\ three\ times$
\STATE $a\ n-gram\ string\ matching\ is\ used\ to\ compare\ two\ fields\ of\ a\ document\ pair$
\STATE $A\ rule\ of\  transitivity\ increases\ a\ cluster:$
	\IF {$R1\ matches\ R2\ and\ R2\ matches\ R3$}
		\STATE $R1\ matches\ R3$
	\ENDIF\end{algorithmic}
\end{algorithm}

A variant of "Algorithm~\ref{alg:AT}", proceeds in two steps, to build a key with bigrams (BGF) (see \cite{Tian:2002}). It has been called anchor method by \cite{Manber:1994}. The first step is the elaboration of a dictionary of bigrams from databases, theoretically, according to the alphabet (see its definition, above) we could find 36x36, hence 1,296 bigrams. After extracting their number of occurrences and excluding ambigous bigrams such as  \textit{at,it, is} ...; we keep the 50 most frequent bigrams constituting the anchor dictionary th, an, re, en, si, er, al, ce, es, pa, di, .... The second step transforms a title according to the present bigrams, from the dictionary, to build a key. For instance, according our example,

\begin{enumerate}
	\item $Title$ = \textit{hybrid inhibitors of phosphatidylinositol 3-pinase (pi3k) and the mammalian target of rapamycin (mtor): design, synthesis, and superior antitumor activity of novel wortmannin-rapamycin conjugates}
	\item $K_{BGF}$ = \textit{id hi bi ph ph id na an he al nt es nt he si an er nt ac vi no ve an ni pa es}
\end{enumerate}

The main concurrent approaches to fingerprinting are more related to definition of a space of descriptors and usage of similarity indices to find the closest semantically related documents. It is more or less based on famous document clustering techniques. In these approaches, as in fingerprinting approaches, heuristics are used to select features and to represent efficiently the content of each document.  \\
Two widespread algorithms are devoted to this approach.   \\
There is the Salton vector space approach (SVS), based on using distance metrics such as Euclidean distance between two documents (or the distance can be something else such as cosine measure or many other measures having the properties of a distance in a high-dimensional vector space cf. \cite{Chowdhury:2002}). In our case, each dimension is associated to a word extracted from the titles and abstracts. The advantage of such a representation is that each dimension can receive a weight according to its presence in a document or the database, hence the famous TF-IDF. This means the product of the number of occurence of a word in the database by the inverse number of occurences in a current document. It weights positively the less frequent and locally-distributed words. The algorithm, "Algorithm~\ref{alg:SVS}"has three different parts. The first one builds a dictionary of words in the whole database and keeps the most frequent ones (more than 2 occurences). The second part splits each document in the same way to make a lexical vector, keeping those already present in the dictionary. It builds a vector in the space of dictionary attributes. The third part chooses each document to classify by comparing it pairwise to each document of the target database. The similarity between such vectors of attributes can be computed using a cosine measure. If the result is greater than a convenient threshold (usually 0.95) then the current document is tagged as a duplicate.  \\
The second approach is nearly the same as the previous one but the attributes are not words but collocations, and especially all 2-collocations and 3-collocations (see \cite{Brin:1995}, \cite{Tamilselvi:2009}). We call it the collocation-similarity-based approach (CSB). As collocations are not so frequent it is more convenient for this approach to use a similarity coefficient to compare the set of collocations of a current document to another one. The algorithm can hence use a famous coefficient such as the Jaccard measure, computing ratio of intersection of attributes between two sets to the union of attributes. "Algorithm~\ref{alg:CSB}" details the routine.
\\

\begin{algorithm}[H]
\caption{Salton Vector Space algorithm}\label{alg:SVS}
\begin{algorithmic}[1]
\medskip
\REQUIRE Databases files \COMMENT {$Dictionary\ building$};
\FOR {$each\ document\ D_{i}\ in\ Databases$ }
	\STATE $select\ fields\ title\ and\ abstract,\ concatenate\ in\ a\ string\ S$
	\STATE $Split\ S\ in\ single\ words$ \COMMENT {example:\ 'mammalian\ target\ of\ rapamycin'\ leads\ to\ the\ set\ {'mammalian','target','of','rapamycin'}}
	\STATE $sort\ list\ of\ words\ and\ store\ each\ word\ in\ a\ hash\ table\ HT\ with\ its\ rank\ r.$
\ENDFOR
\medskip
\REQUIRE A given Document \COMMENT {$Document\ Descriptor\ Building$};
\STATE $select\ fields\ title\ and\ abstract,\ concatenate\ in\ a\ string\ S$
\STATE $Split\ S\ in\ all\ contiguous\ collocations\ of\ size\ 2\ and\ 3,\ store\ in\ LW$
\STATE $Set\ a\ vector\ V\, with $ \#V $ \gets  $ \#HT
	\FOR {$each\ word\ W\ in\ LW\ $}
		\IF {$C\ belongs\ to\ HT,\ with\ rank\ r$}
			\STATE $set\ V[r] \gets 1$
		\ELSE 
			\STATE $set\ V[r] \gets 0$
		\ENDIF
 \ENDFOR					
\medskip
\REQUIRE Descriptor vectors V(i) \COMMENT {$Document\ Comparison\ with\ a\ given\ document\ D_{d}$};
	\STATE $m\ \gets\ $\#HT
	\STATE $Threshold\ \gets\ 0.95$
	\FOR {$each\ vector\ V(c)\ of\ a\ document\ D(c)\ from\ the\ database$}
		\STATE $compute\ a\ cosine\ similarity\ measure$
		\begin{equation} I = \frac{ \sum^{m}_{i=1}(V_{ci}.V_{di}) }{ \sqrt{\sum^{m}_{i=1}V_{ci}^{2}}.\sqrt{\sum^{m}_{i=1}V_{di}^{2}} }  \end{equation} 
	  \IF {$I\ >\ Threshold$}
			\STATE $set\ D(c)\ as\ duplicate\ of\ D(d)$
		\ENDIF
 \ENDFOR					
					
\end{algorithmic}
\end{algorithm}
					
\begin{algorithm}[H]
\caption{Collocation Similarity-Based algorithm}\label{alg:CSB}
\begin{algorithmic}[1]
\medskip
\REQUIRE Databases files \COMMENT {$Dictionary\ building$};
\FOR {$each\ documents\ D_{i}\ in\ Databases$}
	\STATE $select\ fields\ title\ and\ abstract,\ concatenate\ in\ a\ string\ S$
	\STATE $Split\ S\ in\ all\ contiguous\ collocations\ of\ size\ 2\ an\ 3$ \COMMENT {example:\ 'mammalian\ target\ of\ rapamycin'\ leads\ to\ set\ \{'mammalian\ target','mammalian\ target\ of','target\ of',\ 'target\ of\ rapamycin'\}}
	\STATE $sort\ list\ ot\ collocations\ and\ store\ each\ collocation\ in\ a\ hash\ table\ HT\ with\ its\ rank\ r.$
\ENDFOR
\medskip
\REQUIRE A given Document \COMMENT {$same\ routine\ as\ previous\ as\ in\ SVS\ but\ split\ with\ collocations\ -\ Document\ Descriptor\ Building$};	
\medskip
\REQUIRE Descriptor vectors V(i) \COMMENT {$Document\ Comparison\ with\ a\ given\ document\ D_{d}$};
	\STATE $m\ \gets $\#HT
	\STATE $Threshold\ \gets 0.95$
	\FOR {$each\ vector\ Vc\ of\ a\ document\ Dc\ from\ the\ database$}
		\STATE $compute\ a\ Jaccard\ similarity\ measure$
		\begin{eqnarray} J_{DC}=J(V_{D},V_{C})=\frac{ \left| \left\{ V_{D0},...,V_{Dm} \right\} \cap \left\{ V_{C0},...,S_{Cm} \right\} \right| }{ \left| \left\{ V_{D0},...,V_{Dm} \right\} \cup \left\{ V_{C0},...,S_{Ck} \right\} \right| }	\end{eqnarray}
	  \IF {$I\ >\ Threshold$}
			\STATE $set\ D(c)\ as\ duplicate\ of\ D(d)$
		\ENDIF
 \ENDFOR					
					
\end{algorithmic}
\end{algorithm}

Other specific approaches try to investigate the properties of documents with original techniques, such as learning, but the results are not convincing. \cite{Bilenko:2003} proposes to learn the parameters (weights) between matrices of a generative model for string distance. For that they used forward-backward and EM algorithms. For each field they compute a distance measure hence making a vector of different measures they trained a SVM binary classifier to sort duplicates/non-duplicates. The result gives 100\% of recall, precision is around 45\%.

\section[Our Language Models for Fingerprinting]{Our Language Models for Fingerprinting}\label{sec:olmf}

\subsection[Key definition]{Key definition}\label{subsec:kd}

We make two keys from titles and authors which seems to be the most specific to scientific literature. In this way it may be applied to other documents (work documents, emails, blogs, ...) having generally a title and an author. Websites of course have a different document structure but this is not our purpose. 
\\
We decided to use two different methodologies for building a key. One is relatively intuitive, taking into account word-sequence and the author name (socio-semantic approach, or SSF). The second does not take this into account and is only based on compaction using the alphabet (monogram approach MGF), and a variant by sorting the key (sorted monogram approach, SMGF). 
\\
SSF proceeds as follows. First, title and first author name are concatenated into a single string S. Second, S is reduced to lower case. Finally, we keep only the first bigram of each word in the same order of occurrence of the words in the limit of the N first bigrams of the title. Practically, we take N=8. MGF is quite different in that way, the key is obtained only with the help of the alphabet. We check from left to right the first, and only the first, occurrence, keeping the order, of a letter from the alphabet. The key is reduced to lower case and no delimiters are taken into account. In the SMGF variant, the key is sorted alphabetically. In our example we thus obtain

\begin{itemize}
	\item \textit{AU  -  Ayral-Kaloustian, S     Gu, JX     Lucas, J     Cinque, M    Gaydos, C    Zask, A    Chaudhary, I    Wang, JY    Di, L    Young, M    Ruppen, M    Mansour, TS    Gibbons, JJ    Yu, K}
	\item \textit{TI  - Hybrid Inhibitors of Phosphatidylinositol 3-Kinase (PI3K) and the
   Mammalian Target of Rapamycin (mTOR): Design, Synthesis, and Superior
   Antitumor Activity of Novel Wortmannin-Rapamycin Conjugates}
	\item $K_{SSF}$  = \textit{hy in of ph 3- (p an th ma ta ay}
	\item $K_{MGF}$  = \textit{hybridntorspal3keumcvwjg}
	\item $K_{SMGF}$ = \textit{abcdeghijklmnoprstuvwy3}
\end{itemize}

These techniques achieve unique key extraction with a very compact representation using only 20 characters on average.

\subsection[Algorithm]{Algorithm}\label{subsec:al}

The algorithm has been written in Perl, and is quite fast. If the size of the test database is N and the size of the target database is M we need to compute a key for each (N+M) documents. This step has complexity O(N+M) in time. The second step relies upon hash lookup. Lookup operations in a 'map' data structure are guaranteed to have a time complexity that is at most logarithmic in the number of key-value pairs in the map. Hence for each test document we need a lookup of N*log( M ). Globally, the time complexity is O( (N+M) + N.log(M) ).  \\ 
Complexity in space is not high. It requires the storage of (N+M) documents, (N+M) pointers to keys and hashkeys, (N+M) lists of words for titles (and author names). So the global complexity is O(N+M). On a laptop with a pentium M processor and 2.3 GhZ clock, N=7,709, M=12,658, the running time is 7 seconds. Space consumed is 254 Mb, around 5 times the size of the databases (the size on disk of the databases is 43 Mb ; 24 Mb for about N documents, 19 Mb for about M documents).\\
The algorithm is developed in two steps. A first step computes a key for each document (see previous chapter for details); a document can belong to the test or target database. The second step picks a document from the test database and its associated key, to compare it to all other keys for the target database. "Algorithm \ref{alg:SSF}" describes the SSF approach. "Algorithm \ref{alg:MGF}" describes the MGF approach.  "Algorithm \ref{alg:SMGF}" describes the SMGF approach which varies a little bit from the MGF approach only by sorting the key.

\begin{algorithm}[H]
\caption{Socio-Semantic Fingerprinting algorithm}\label{alg:SSF}
\begin{algorithmic}[1]
\medskip
\REQUIRE Databases files \COMMENT {$1st\ step,\ keys\ building$};
	\STATE $N\ \gets 8$
	\FOR {$each\ document\ D\ from\ the\ databases$}
		\STATE $select\ title\ (T)\ and\ first\ author\ name\ (A)$
		\STATE $transform\ T\ and\ A\ in\ lower\ case$
		\STATE $split\ T\ into\ words,\ store\ in\ LW$
		\FOR {$each\ words\ W_{i}\ from\ the\ LW,\ and\ i<=N$}
			\STATE $select\ the\ first\ bigram\ of\ W,\ store\ in\ K$
		\ENDFOR
		\STATE $select\ the\ first\ bigram\ of\ A,\ store\ in\ K$					
 \ENDFOR					
\STATE $K\ is\ stored\ in\ a\ hash\ table\ HT$
\medskip
\REQUIRE given hash key of test document D(c), hash table HT of target databases \COMMENT {$2nd\ step,\ comparison$};
	\IF {HT ($K_{D(c)})\ return\ d$}
		\STATE $set\ D(c)\ as\ duplicate\ of\ D(d)$
	\ENDIF					
\end{algorithmic}
\end{algorithm}

\begin{algorithm}[H]
\caption{Monogram Fingerprinting algorithm}\label{alg:MGF}
\begin{algorithmic}[1]
\medskip
\REQUIRE Databases files, Alphabet A \COMMENT {$1st\ step,\ keys\ building$}
	\STATE $Miror_LA\ \gets $\#A 
	\FOR {$each\ document\ D\ from\ the\ databases$}
		\STATE $select\ title\ (T)$
		\STATE $transform\ T\ in\ lower\ case$
		\STATE $split\ T\ into\ letters\ of\ A,\ store\ in\ LA$
		\FOR {$each\ letter\ L_{i}\ from\ LA$}
			\IF {$Miror_LA(L_{i})\ <>\ 0$}
				\STATE $store\ L_{i} in K$
				\STATE $Miror_LA(L_{i})\ \gets 1$
			\ENDIF					
		\ENDFOR
 \ENDFOR					
\STATE $K\ is\ stored\ in\ a\ hash\ table\ HT$
\medskip
\REQUIRE   D(c),  HT   \COMMENT {$2nd\ step,\ same\ as\ $ "algorithm \ref{alg:SSF}" }
\end{algorithmic}
\end{algorithm}

\begin{algorithm}[H]
\caption{Sorted Monogram Fingerprinting algorithm}\label{alg:SMGF}
\begin{algorithmic}[1]
\medskip
\REQUIRE Databases files, Alphabet A \COMMENT {$1st\ step,\ keys\ building$}
\STATE $\ same\ as\ $ "algorithm \ref{alg:MGF}"
\STATE $\ sort\ K\ by\ alphabetical\ order$
\STATE $K\ is\ stored\ in\ a\ hash\ table\ HT$
\medskip
\REQUIRE   D(c),  HT   \COMMENT {$2nd\ step,\ same\ as\ $  "algorithm \ref{alg:MGF}"}
\end{algorithmic}
\end{algorithm}

\section[Evaluation]{Evaluation}\label{sec:ev}

Two protocols were implemented, one using benchmark data (i.e., a gold standard), the extrinsic evaluation; the other using only datasets, the intrinsic evaluation.

\subsection[Gold standard data]{Gold standard data}\label{subsec:gsd}

An offline step has been achieved to characterize a test set defined by hand. In this step we chose a set of documents in which we search for an exact duplicate. We fixed the size of this set to be 374 documents from the PM dataset. In this set, 202 documents were found to be exact duplicates. They settle a gold standard for future comparisons.

\subsection[Intrinsic evaluation]{Intrinsic evaluation}\label{subsec:ie}

The intrinsic evaluation is achieved within 3 protocols.\\
We call the first one, the Monte Carlo approach (MC). The test database is the gold standard set of 374 documents. No target database is defined and a tag is randomly chosen. The results are shown in Figure  \ref{figure2}, almost 50\% of precision and recall. It constitutes a kind of baseline for the rest of evaluation.\\
The  second and third protocols use only PM database. In the second protocol, the PM database is split into two separate files with no redunduncy; we call this protocal HPM (half PubMed evaluation). Theoretically the method should not find any duplicates. In practice it makes one mistake among the 3,855 documents. It finds that the following document, resembling a quasi-plagiat by the author, is a quasi-duplicate: 

\begin{enumerate}
	\item WOS sentence :  \textsl{TI  - Clonal relationship among Vibrio cholerae O1 El Tor strains isolated in Somalia.
AB  - One hundred and three Vibrio cholerae O1 strains, selected to represent the cholera outbreaks which occurred in Somalia in 1998-1999, were characterized by random amplified polymorphic DNA patterns, ribotyping, and antimicrobial susceptibility...
AD  - Dipartimento di Genetica e Microbiologia, Universita di Bari, Italy.
FAU - Scrascia, Maria} (PMID- 18774337)
	\item PM sentence : \textsl{TI   Clonal relationship among Vibrio cholerae O1 El Tor strains causing the largest cholera epidemic in Kenya in the late 1990s.
AB   Eighty Vibrio cholerae O1 strains selected to represent the 1998-to-1999 history  of the largest cholera epidemic in Kenya were characterized by ribotyping, antimicrobial susceptibility, and random amplified polymorphic DNA patterns....
AU - Scrascia, Maria} (PMID- 16954285)
\end{enumerate}

The last intrinsic evaluation (third protocol) concerns usage of the PM database as test and target at the same time. We call this protocol FPM (Full PubMed evaluation). Theoretically it has to find all the documents as duplicates, and it does. So, the approach gets a rate of success quite reasonable at this step of evaluation.\\

\begin{figure}[!ht]
\centering
\renewcommand{\arraystretch}{1.2}
{\scriptsize
\begin{tabularx}{15.5cm}{|>{\centering\hsize=0.4\hsize\arraybackslash}X|>{\centering\hsize=0.7\hsize\arraybackslash}X|>{\centering\hsize=0.7\hsize\arraybackslash}X|>{\centering\hsize=0.5\hsize\arraybackslash}X|>{\centering\hsize=0.6\hsize\arraybackslash}X|>{\centering\hsize=0.8\hsize\arraybackslash}X|>{\centering\arraybackslash}X|>{\centering\hsize=0.5\hsize\arraybackslash}X|}
   \hline
   \textbf{  } & \textbf{ \#Gold Duplicates } & \textbf{\#Predicted duplicates} & \textbf{ \#True positives} & \textbf{ \#False positives } & \textbf{ \#False negatives } & \textbf{ Precision ratio } & \textbf{ Recall ratio }  \\
   \hline
   \textbf{MC} & 202 & 203 & 94 & 109 & 108 & 0.463 & 0.465 \\
\hline
   \textbf{HPM} & 0 & 1 & 0 & 1 & 3855 & 1.00 & 1.00 \\
\hline
   \textbf{FPM} & 7709 & 7709 & 7709 & 0 & 0 & 1.00 & 1.00 \\  
\hline
\end{tabularx} } 
\caption{Comparison between methods with a gold standard.}\label{figure2}
\end{figure}

\subsection[Comparison with other methodologies]{Comparison with other methodologies}\label{subsec:com}

In this benchmark, we tested 10 methodologies to compare documents. The initial goeal is to compare pairwise documents between two different database, specifically PubMed and WebofScience to merge two corpora and delete duplicates. \\
Our techniques presented here are based on very compact language models to make a unique key for each document: socio-semantic fingerprint (SSF,  "Algorithm \ref{alg:SSF}"), monogram fingerprint (MGF,  "Algorithm \ref{alg:MGF}"), and the sorted monogram fingerprint (SMGF,  "Algorithm \ref{alg:SMGF}" ). Three approaches from the state of the art: the Salton vector space (SVS,  "Algorithm~\ref{alg:SVS}"), collocation similarity-based (CSB,  "Algorithm~\ref{alg:CSB}") and the bigram fingerprint (BGF,  "Algorithm~\ref{alg:AT}"). Moreover, three simple key fingerprint approaches also were tested: author fingerprint (AF), title fingerprint (TF), author-review-date fingerprint (ARDF), and the modified title fingerprint (MTF) ("Algorithm~\ref{alg:SN}").\\
We have  three summarising scores. The basic methodologies AF, TF, ARDF and MTF are not efficient with recall score because we prefer recall maximization. SVS, CSB and BGF give good recall but precision is not so good as SSF and MGF. SMGF is catastrophic, pointing out the importance of the sequence order of the letters in a string. SVS has good results but unfortunately many more false positives than SSF. CSB differs from SSF only by 1 false positive in the CSB side. Nevertheless we can assert that SSF is at the top of the results, with MGF having no false positives and 2 false negatives less than SSF.

\begin{figure}[!ht]
\centering
\renewcommand{\arraystretch}{1.2}
{\scriptsize
\begin{tabularx}{15.5cm}{|>{\centering\hsize=0.4\hsize\arraybackslash}X|>{\centering\hsize=0.7\hsize\arraybackslash}X|>{\centering\hsize=0.7\hsize\arraybackslash}X|>{\centering\hsize=0.5\hsize\arraybackslash}X|>{\centering\hsize=0.6\hsize\arraybackslash}X|>{\centering\hsize=0.8\hsize\arraybackslash}X|>{\centering\arraybackslash}X|>{\centering\hsize=0.5\hsize\arraybackslash}X|}
   \hline
   \textbf{  } & \textbf{ \#Gold Duplicates } & \textbf{\#Predicted duplicates} & \textbf{ \#True positives} & \textbf{ \#False positives } & \textbf{ \#False negatives } & \textbf{ Precision ratio } & \textbf{ Recall ratio }  \\
   \hline
   \textbf{SSF} & 202 & 192 & 192 & 0 & 10 & 1.00 & 0.95 \\	 	 	 
   \hline
   \textbf{TF} & 202 & 131 & 131 & 0 & 71 & 1.00 & 0.649 \\
   \hline
   \textbf{MTF} & 202 & 151 & 151 & 0 & 51 & 1.00 & 0.748 \\
   \hline
   \textbf{AF} & 202 & 314 & 202 & 112 & 0 & 0.643 & 1.000 \\
   \hline
   \textbf{ARDF} & 202 & 139 & 134 & 5 & 68 & 0.964 & 0.663 \\
\hline
   \textbf{MGF} & 202 & 194 & 194 & 0 & 8 & 1.00 & 0.960 \\
\hline
   \textbf{SMGF} & 202 & 295 & 197 & 98 & 5 & 0.668 & 0.975 \\
\hline
   \textbf{BGF} & 202 & 196 & 193 & 3 & 9 & 0.985 & 0.955 \\
\hline
   \textbf{SVS} & 202 & 212 & 201 & 11 & 1 & 0.948 & 0.995 \\
\hline
   \textbf{CSB} & 202 & 193 & 192 & 1 & 10 & 0.995 & 0.950 \\
\hline
\end{tabularx} } 
\caption{Comparison between methods with a gold standard.}
\refstepcounter{figure}\label{figure3}\addtocounter{figure}{-1}
\end{figure}

\section[Conclusion]{Conclusion}\label{sec:conc}

Duplication has been presented in this paper as a modern topic having applications such as merging databases or plagiarization detection. We aimed at merging two bibliographic record databases. In this area, cleaning is an important process, to eliminate redundant information. We presented several interesting and efficient algorithms from the state of the art. These algorithms can be improved.\\ 
Our language models, divided into two fingerprint key extraction, socio-semantic, and monogram keys, are able to get good scores in evaluation protocols compared to other algorithms. We get more than 95\% recall with 100\% precision.

\section[Acknowledgments]{Acknowledgments}\label{sec:ack}
Special thanks to Prof. Dr. Christian Meyer, at INRA Paris, a specialist in molecular biology and plant biology for cooperation about the TOR molecule.

\nocite{*}
\bibliographystyle{plain}
\bibliography{TurenneDUP}
\end{document}